\documentclass[conference]{IEEEtran}

\usepackage{graphicx}  

\usepackage{subfigure}

\begin{document}

\title{{\small [Fortieth Hawaii International Conference on System Sciences, January 3-6, 2007, Big Island, Hawaii]}\\
        Stochastic Model for  Power Grid Dynamics}

\author{\authorblockN{Marian Anghel}
\authorblockA{Los Alamos National Laboratory\\
Los Alamos, New Mexico 87544\\
Email: manghel@lanl.gov}
\and
\authorblockN{Kenneth A. Werley}
\authorblockA{Los Alamos National Laboratory\\
Los Alamos, New Mexico 87544\\
Email: kaw@lanl.gov}
\and
\authorblockN{Adilson E. Motter}
\authorblockA{Northwestern University\\
Evanston, Illinois 60208\\
Email: motter@northwestern.edu}}

\maketitle

\begin{abstract}
  We introduce a stochastic model that describes the quasi-static
  dynamics of an electric transmission network under perturbations
  introduced by random load fluctuations, random removing of system
  components from service, random repair times for the failed
  components, and random response times to implement optimal system
  corrections for removing line overloads in a damaged or stressed
  transmission network. We use a linear approximation to the network
  flow equations and apply linear programming techniques that optimize
  the dispatching of generators and loads in order to eliminate the
  network overloads associated with a damaged system.  We also provide
  a simple model for the operator's response to various contingency
  events that is not always optimal due to either failure of the state
  estimation system or due to the incorrect subjective assessment of
  the severity associated with these events. This further allows us to
  use a game theoretic framework for casting the optimization of the
  operator's response into the choice of the optimal strategy which
  minimizes the operating cost. We use a simple strategy space which
  is the degree of tolerance to line overloads and which is an
  automatic control (optimization) parameter that can be adjusted to
  trade off automatic load shed without propagating cascades versus
  reduced load shed and an increased risk of propagating cascades.
  The tolerance parameter is chosen to describes a smooth transition
  from a risk averse to a risk taken strategy. We present numerical
  results comparing the responses of two power grid systems to
  optimization approaches with different factors of risk and select
  the best blackout controlling parameter.
\end{abstract}

PACS: 89.75.-k, 05.10.-a, 02.50.-r

\IEEEpeerreviewmaketitle

\section{Introduction}

It is well-known that power grids are among  the largest and most complex 
technological systems ever developed. These systems suffer periodic disturbances at 
various scales and some times these disturbances are so large to affect
a considerable fraction of the power grid and induce considerable economic and social
costs. For this reason, the vulnerability of power grids and more generally
of interconnected transportation and communication networks has been
recently studied extensively. 
In order to understand the global dynamics of power system blackouts a
computationally efficient \emph{global system analysis} approach is
required in order to capture the overall response of the system to
such events~\cite{Ca01,Do04}.  Moreover, because this approach should
be able to describe the random sequences of rare events that trigger
large blackouts, it should have a stochastic character.  On the other hand,
 any probabilistic risk assessment requires a large number of long
simulations in order to determine the thought after risk indexes.
Unfortunately, because the probability of these events is very low,
it is very difficult to model the full complexity of the network's
dynamics over the very long time scales that are necessary for
describing these rare events. 
One way to approach this difficult problem is to include simple, but
representative, models for each component of this large and complex
system as 
required in order to understand the global dynamics of
power system blackouts.  This is the approach followed by Dobson and
coworkers~\cite{Do01,Ca01,Ca02,Ca04}, who introduced a model
that does not attempt to simulate the complex details of each blackout
event, which may be comprised of complicated processes involving
protection systems, dynamics, and human factors.  Instead, the
blackout cascades in this model are essentially instantaneous events
due to dynamical redistribution of power flows and are triggered by
probabilistic failures of overloaded lines.  The size of blackouts is
determined by solving a standard LP optimization of the generation
dispatch, consistent with the power flow equations and operational
constraints, and the redistribution of power flows is calculated using
a linear load flow approximation.

Thorp and his collaborators ~\cite{Th98,Ba99,Ch02} perform a
reliability study of transmission system protection devices using
linear load flow approximation and linear programing optimization of
generation dispatch and load shedding operations.  Their approach uses
a probabilistic model to simulate the incorrect tripping (sympathetic
tripping) of lines and generators due to hidden failures of line or
generator protection systems.  The distribution of power system
blackout size is obtained using importance sampling and blackout risk
assessment and mitigation methods are also studied.

In another model, Rios \emph{et.~al.~}~\cite{Ri02} propose a
 probabilistic cost/benefit analysis and calculate the expected blackout cost,
which they equate with the value of grid security, using Monte Carlo
simulations of a power system model that represents the effects of
cascading line overloads, hidden failures of the protection system,
and power system dynamic instabilities. In this approach the authors replace the complex
time-dependent modeling of transient instability with a sophisticated
probabilistic modeling using off-line computation of conditional
probabilities based on the fault type and its location with respect to
the vulnerability region of each generator.

Another example of advanced analysis of cascading failures and
high-consequence contingencies, performed using TRELSS software, is
described by Hardiman \emph{et.~al.~}\cite{Ha04}.  In its simulation
approach mode, TRELSS simulates system vulnerability to cascading
failures which are initiated by outages of lines, transformers, and
generators due to overloads and voltage violations. The nodal voltages
are monitored and if any load bus voltage magnitude becomes less than
a specified ``voltage collapse'' threshold, the corresponding load is
dropped. Similarly, voltages are controlled at generator buses as well
and if they become less than a specified threshold, the corresponding
generators are also tripped.  Note though, the different conceptual
approach to estimating voltage dependent effects, compared to the
probabilistic approach of Rios \emph{et.~al.~}~\cite{Ri02}.

Ni \emph{et.~al.~}\cite{Ni03} provide on-line risk-based security
assessment (OL-RBSA) for rapid quantification of the security level
associated with an existing or forecasted operating condition.  Their
probabilistic approach condenses contingency likelihood and severity
functions into indexes that reflect probability risks. These indexes
include overload security (flow violations and cascading overloads)
and voltage security (voltage magnitude violations and voltage
instability), while the risk assessment involves both the modeling of
uncertainty as well as the modeling of severity functions assuming no
operator's action (such as redispatch) occurs. The goal is to reflect
the consequence of the contingency and loading conditions, rather than
the consequences of an operator's decision.

These studies
emphasize the importance of building representative
stochastic models for the global system analysis of network
reliability and of cascading failure risk.
It is surprising that 
in most previous works,
the modeling of the operator's response to 
these
complex and
often unpredictable contingencies is very simple and, paradoxically,
predictable. Because it is difficult to manage the technical difficulties
associated with low-probability, high-impact events, modeling the
often imperfect operator's response should be an important component of
the network reliability and blackout risk assessment. To emphasize this
fact, we point out that the systematic control of large power systems
in response to major contingency events is effectively nonexistent.
The methods used are expert-based and are by rule not automated.
Therefore, the inclusion of a human decision maker is
critical \cite{IlSk00}.
Moreover, because most models generate only abstract blackouts, as
 sequences of technical failures propagating in space but
 instantaneous in time, they lack a faithful representation of the
 evolution of these events in time.  This is unfortunate because it
 prevents us from using the sequence of system fluctuations, produced
 during the often slow, initiating phase that precedes large cascading
 events, to perform sophisticated pattern analyses aimed at
 predicting a developing blackout event as an anomaly detection
 problem.

Here we describe our first steps into the implementation of a power
grid dynamic model that incorporates simple stochastic rules for the
reliability of each grid component and offers a better description of
line outages and line restoration events using a simple time-dependent 
approach.
Unlike previous models, this model describes the utility response to
various disturbances in an attempt to include a description of the
human response to contingency events that is not always optimal due to
either failure of the state estimation system or due to the incorrect
subjective assessment of the severity associated with these events.
In particular, this model is expected to serve as a general framework
for a realist system-level analysis of the power grid from the viewpoint of
the theory of complex networks \cite{n1,n2,n3,n4,n5,n6,n7,n8}.

\section{Stochastic Grid Model}
\label{grid_model}

Our approach, inspired by the model introduced
in~\cite{Do01,Ca01,Ca02,Ca04}, computes the distribution of
power flows using a linear load flow approximation, includes optimal
generator dispatch and load shedding operations in order to alleviate
operating contingencies, and models blackouts by overloads and outages
of transmission lines. Disturbances in the grid are introduced by line
outages due to unforeseen stochastic events. In order to describe the
evolution of cascading events in their slow initiating stages,
transmission lines failures in our model are due to line overheating
due to excessive power flows. To describe this effect we monitor the
evolution of the line temperature, and its slow dependence on flow
redistributions, using as a model the conduction of heat in rods of
small cross-section in which an electric current of constant strength
is flowing. Further contributing to the slow evolution of cascades, is
a line restoration model which prevents a damaged line from being put 
back in service before a random restoration time has passed.  The model
also has the ability to follow daily and seasonal changes in demand,
although
these features have not been used in the simple optimization
experiments presented in this paper. 
We also provide a bounded
rationality model for the operator's response that is some times
suboptimal due to either failure of the data acquisition equipment or
due to the incorrect subjective assessment of the severity associated
with cascading overloads. This further allows us to use a game
theoretic framework for casting the operator's response optimization
into the choice of the optimal strategy which minimizes the operating
cost.  In the stochastic model introduced here the strategy space is
defined by a line overload parameter which describes a smooth
transition from a risk averse to a risk taken response.

The model and simulations were performed within the framework offered
by the Power System Analyzer (PSA)~\cite{We05}, which is a suite of
numerical tools developed at Los Alamos National Laboratory to permit
model building, simulation, analysis, and graphical display of
electric power transmission networks.  Before presenting the details
of the simulation algorithm we describe in detail each of the
components of the stochastic model.

\subsection{Formulation of the DC power flow}

We assume that the electrical transmission system operates in
steady-state conditions and that this assumption holds even during 
the evolution of major disturbances in the system.  This is obviously
an approximation which is violated during the late stages of major
disturbance events, but the model can be modified to better describe
these events by modeling voltage-dependent phenomena.

In order to determine the steady-state operating conditions of the
power grid, we should solve the full nonlinear power flow equations
that provide information about the voltage magnitudes and phases and
the active and reactive power flows along each transmission line.
Unfortunately, since our simulations involve numerous power flow
solutions for a power grid system that evolves in time under various
random perturbations, solving repeatedly the full nonlinear power
flow equations becomes computationally prohibitive. Moreover, we are
interested in estimating the statistical properties of various
quantities that characterize the system's response to these
perturbations, like the frequency distribution of blackout sizes,
their duration, the distribution of inter-event times, etc., which also
require the simulation of tens of thousands of blackout events.  
In addition, we are interested in estimating the optimal strategy necessary
to mitigate the impact of blackout events. To address this problem, the
full nonlinear equations pose very difficult nonlinear optimization
problems. For all these reasons, we have chosen to linearize the
power-flow equations and to solve instead the so called ``DC'' power
flow equations that connect the flow of real power to the voltage
phases of the system's buses, which results in a completely linear,
non-iterative power flow algorithm~\cite{WoWo96}. 

The DC power flow can only calculate real (MW) flows on transmission
lines and 
gives no answers to what happens to voltage magnitudes or
reactive (MVAR) flows. Assuming that all bus voltages phasors are 1.0
per unit in magnitude, and defining the matrix $ \mathbf{B}$ by
$B_{ij} = - b_{ij}$ if $i\neq j$ and $B_{ii} = \sum_{j\neq i} b_{ij}$, where $b_{ij}$ is
the susceptance of the transmission line joining buses $i$ and $j$,
the voltage phases $\theta_i$ are the solution of the linear power flow equation
$\mathbf{P} = \mathbf{B} \mathbf{\Theta}$. 
Here, $\mathbf{P}$ is the
vector whose $N-1$ components are the real powers injected at each
node, except for a reference node (slack node) for which the injected real
power is computed from the power balance between total generation and
total load.  The vector $\mathbf{\Theta}$ is the vector whose 
components are the voltage phases at each node in the network except
the slack node which has phase zero. After solving the power flow
equation for the vector $\mathbf{\Theta}$, the flow of real power
along each transmission line is computed from $P_{ij} =
b_{ij}(\theta_i - \theta_j)$.

\subsection{Random line failure model}

We assume that cascades in the grid are triggered by random line
failure events and that lines fail independently of each other. For a
line $l$, its random failure events are described by a Poisson process
of constant rate $\lambda_{fl}$ such that the number of events in any
interval of length $t$ 
follows
a Poisson distribution with mean
$\lambda_{fl} t $. Hence,  the number of events
which have occurred along line $l$ up to time $t$,  $N_l(t)$, has the following
probability distribution:
\begin{equation}
P_l\{N_l(t) = n\} = e^{-\lambda_{fl} t}\frac{(\lambda_{fl} t)^n}{n!}\, , 
          \hspace{0.5cm} n \ge 0\, ,
\end{equation}
while its expectation is given by
\begin{equation}
E_{P_l}[ N_l(t)] = \lambda_{fl} t\, ,
\end{equation}
which shows that $\lambda_{fl}$ is the average density of failure points.
Assuming a constant reference rate per unit length, $\lambda_f$, for
all lines in the network, the failure rate $\lambda_{fl}$ scales
proportionally with the line length $L_l$ and is given by
$\lambda_{fl} = \lambda_f L_l$.

In order to generate the random failure points $t_n$, $n=1,2,\ldots$,
 we sample
from the stochastic process of rate
 $\lambda_{fl}$.  The sampling process is very simple because the
 interval $x = t_n - t_{n-1}$ between two consecutive failure points
 has an exponential distribution of density~\cite{Pa84},
\begin{equation}
f_l(x) = \lambda_{fl} e^{-\lambda_{fl} x}\, . \label{exp_distrib}
\end{equation}
The first point in the sequence, $t_1$, is sampled using the fact
that its random distance from the starting point of our
simulation, $t_0$, is described by the same exponential distribution
given by Eq.~(\ref{exp_distrib}).
Therefore, if $\xi_0$ is
sampled from 
distribution (\ref{exp_distrib}) then $t_1 = t_0 +
\xi_0$, and if $\xi_1$ is another number sampled from the same
distribution then $t_2 = t_1 + \xi_1$, and so on. Sampling from an
exponential distribution is 
numerically efficient, and is described in~\cite{numrec}.

The failure rate used in our simulations, $\lambda= 0.0001$, has not
been benchmarked to utility experience.  
Our model
assumes that failure events are independent in space and time but
these are approximations that can be relaxed.
For example, as we have already pointed out, hidden
failures in the protection system can cause intact equipment to be
unnecessarily removed from service (sympathetic tripping) following a
fault on a neighboring component. Moreover, a component that failed
yesterday is more likely to fail in a similar way the next day. One
way to include these temporal correlations is to introduce a ``hidden
failure'' (HF) state to each protection device and a Markov model
describing the transition between the ``ON'', ``OFF'' and ``HF''
states of the device as suggested in~\cite{Pe01}.  Similarly, weather
effects (storms, hot weather, winds, etc) can induce correlated
failures in space. These effects can be easily included in our model
due to the dependence of the overloaded-line failure model on the
ambient temperature and wind velocity. Depending on the weather
conditions to which they are exposed, changes in the failure rates
$\lambda_{fl}$ can also be introduced as described in ~\cite{Ri02}.

\subsection{Overloaded-line failure model}

In order to model the failure of transmission lines due to  loading
over their transmission capacity, we consider the problem of
conduction of heat in rods of small cross-section~\cite{CaJa59} in
which an electric current of constant strength is flowing.  We assume
for simplicity that the transmission line is so thin that the
temperature at all points of its cross-section is the same. We suppose
that the transmission line has constant area of cross-section
$\omega$, perimeter $p$, thermal conductivity $K$, electrical
conductivity $\sigma$, density $\rho$, specific heat $c$, and diffusivity
$\kappa$. We further assume that the heat flux across the surface of
the line is proportional to the temperature difference between the
surface and the surrounding medium and is given by $H(T-T_0)$, where
$T$ is the temperature of the line, $T_0$ is the temperature of the
medium, and $H$ is the surface conductance. The problem of heat
conduction then becomes one of linear heat flow in which the
temperature is specified by the time $t$ and the 
position
$x$ measured
along the transmission line. Indeed, balancing the total rate of gain
of heat in an element of volume bounded by the cross-sections at $x$
and $x+dx$ to the rate at which it gains heat across these faces minus
the heat lost at the surface, we find the following heat
equation,

\begin{equation}
\frac{\partial T(x,t)}{\partial t} = \kappa \frac{\partial^2T(x,t)}{\partial x^2} 
+ \alpha I^2 -  \nu (T(x,t)-T_0)\, ,  \label{temp_eq1}
\end{equation}
\noindent where $\nu =Hp/(\rho c \omega)$, $\alpha = 0.239 /(\rho c
 \omega^2 \sigma)$, $ \kappa = K/(\rho c)$ and $I=P/V$ is the current
 in the line measured in amperes.

In order to estimate the surface conductance $H$ we will assume that 
 the loss of heat across the surface of the
line  is due to forced convection.  When fluid (gas or liquid) at temperature $T_0$ is forced
rapidly past the surface of the line, it is found experimentally that
the rate of loss of heat from the surface is given by $H(T-T_0)$, with a value
of the coefficient $H$ that depends on the velocity and the nature of
the fluid and the shape of the surface~\cite{CaJa59}.  For turbulent flow of air
with velocity $u$ perpendicular to a circular cylinder of diameter
$d$, $H=8 \times 10^{-5} (u/d)^{1/2}$ $\textstyle{cal/(cm^2 s\,
K)}$.

Assuming that fluctuations in power flows along the transmission lines
propagates much faster than any heat flow transients, and since the
heat source is equally distributed along the line, we can neglect the
spatial variation in temperature along the line in order to get 
a
simple equation describing the time evolution in the temperature of the line 
in term of the power flowing through the line:

\begin{equation}
\frac{\partial T(x,t)}{\partial t} = \alpha I^2 -  \nu (T-T_0)\, . \label{temp_eq2}
\end{equation}

If the line is initially at temperature $T(0)$ and the
power flowing through the line has the constant value $P$, the line temperature 
evolves according to this simple equation:

\begin{equation}
T(t) = e^{-\nu t}(T(0)-T_e(P)) + T_e(P)\, , \label{temp_evol}
\end{equation}

\noindent where 
\begin{equation}
T_e(P) = \frac{\alpha}{\nu}\frac{P^2}{V^2} + T_0 \, 
\end{equation}
is the equilibrium temperature that the line will reach when $ t
\rightarrow \infty$.  If at some moment the power flow changes, we
reset the clock and the initial temperature and use the same equation
to describe the evolution of line temperature starting from this
moment on.

A transmission line failure due to excessive heating, followed by line
sagging and tripping, will happen if the present power flow through
the line exceeds the maximum line rating. For each line $l$, we denote
by $T_{cl}$ the equilibrium temperature corresponding to a constant
power flow equal to the line rating $P^{max}_l$, i.~e.~ $T_{cl} =
T_e(P^{max}_l)$. When the power flow through the line changes such
that the 
new power flow $P_l$ exceeds $P^{max}_l$, the line will
start heating toward the new equilibrium temperature.
Since this equilibrium temperature exceeds $T_{cl}$,
at some time $t_{cl}$ the line temperature will reach $T_{cl}$ and the
line will fail. The failure time $t_{cl}$, measured from the moment
when the grid topology and the line flow has changed, can be easily
deduced from Eq.~(\ref{temp_evol}) and is given by
\begin{equation}
t_{cl} = \frac{1}{\nu}\ln \frac{T_{cl} - T_e(P_l)}{T(0)-T_e(P_l)}\, .
\end{equation}

Let us further remark that the line rating, $P^{max}$, and some
reference values for $u$ and $T_0$ are chosen to determine the
critical temperature of the line. This means that on colder days, when
$T'_0 < T_0$, the line will reach its critical, failure temperature,
at a larger power flow $P' > P^{max}$.  Indeed, from the equation
$T_e(P') = T_c$ we get,
\begin{equation}
P' = P^{max} \sqrt{ 1 + (T_0-T_0') \frac{\nu}{\alpha} \left(\frac{V}{P^{max}} \right)^2 }\, .
\end{equation}
Similarly, on a windier day, when $u' > u$ and therefore $\nu' > \nu$,
the line fails when the power flow 
reaches a larger value $P'$ given by
\begin{equation}
P' = P^{max} \sqrt{\frac{\nu'}{\nu}}\, .
\end{equation}
It will be interesting to study how these weather effects, and the associated
fluctuations in wind and temperature across a very large grid, will
impact the risk of large blackout events. As we remarked before, these changes
induce spatial correlation in the  response of transmission lines 
to changes in power flows and, when overloaded,  in the estimated failure times.

\begin{figure*}
\centerline{\subfigure[Line I]{\includegraphics[width=3.9in,height=2.1in]{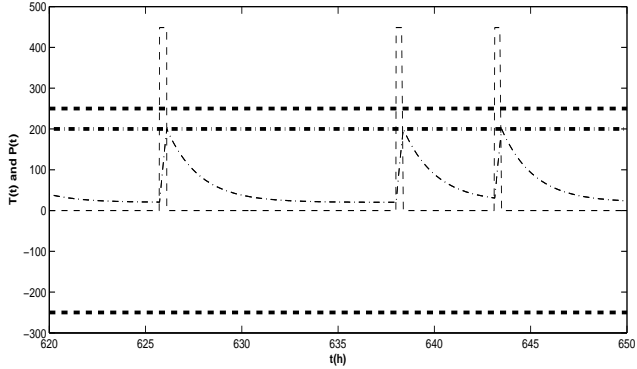}
\label{TP113}}
\hspace{-0.8cm}
\subfigure[Line II]{\includegraphics[width=3.9in,height=2.1in]{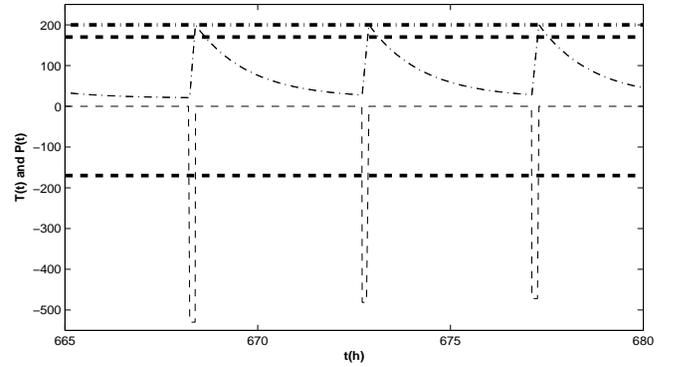}
\label{TP66}}}
\caption{The evolution of power flow (dashed line) and temperature (dash-dotted line)
for two transmission lines in a simulation example. The time is
measured in hours. The horizontal dash-dotted lines represents the critical
temperatures at which each line failure due to overheating happens and
the horizontal dashed 
lines represent the power line ratings.  Both figures
show the repeated failure of the lines shortly after a restoration
event. Since the network has not fully recovered from the initial
failures in other points of the network, each line restoration produces overflows which will shortly
induce another overheating failure event. These examples were chosen to
show that a suboptimal restoration strategy produces only temporary 
relief for a stressed system.}
\label{TPflow}
\end{figure*}

Finally, in order to keep the heat equation linear, we have omitted on
the right hand side of Eq.~(\ref{temp_eq1}) a cooling term that takes
into account that each element of the surface of the rod loses heat
by radiation to the surrounding medium --- and provides cooling when the
wind is absent.  When this term is included, the right hand side of
Eq.~(\ref{temp_eq2}) acquires an additional cooling term
$\frac{H'p}{\rho c \omega} (T^4-T_0^4)$, where $H'$ is a
constant. When the heating of the line is small and $T \approx T_0$,
we can linearize this term and recover Eq.~(\ref{temp_eq2}) with a
corrected constant $\nu$ due to the fact that $H$ is now replaced with
$ H+4H'$.  When the heating of the line is large, we cannot make this
approximation, the equation becomes nonlinear, and has to be
integrated numerically. We have chosen not to do this, but we can
easily include this effect at a reasonable computational cost.

Figure~\ref{TPflow} shows a fragment from the evolution of the power
 flows  and temperatures for two transmission lines
 during the evolution of a power grid subjected to random failure
 events.  The dash-dotted  lines represent the critical temperatures and
 the dashed  lines the line ratings. The time segments were taken
 after increased line flows due to random component failures somewhere
 else in the network have induced the failure of each line due to
 overheating. Both figures show the repeated failure of the lines
 shortly after a restoration period. Since the network has not fully
 recovered from the initial failures, each line restoration produces
 overflows which will shortly induce another overheating failure
 event. In Figure \ref{TPflow} also we notice that due to thermal inertia
 effects the temperature evolution smoothes out the rapid power flow
 fluctuations.

\subsection{Line restoration model}

We choose a very simple line restoration model that assumes that the
restoration time $t_r$ has a constant component, $c$, plus a random variable 
 $w$ that has an exponential distribution
with parameter
$\lambda_r$, i.~e.~
\begin{equation}
t_r = c + w\, , \label{restoration_time}
\end{equation}
where $w$ is sampled from the 
following exponential distribution:
\begin{equation}
f(w) = \lambda_r e^{-\lambda_r w}\, . \label{restoration_pdf}
\end{equation}
\noindent We further assume that $\lambda_r$ is the same for all lines
in the system. The constant $c$ introduces a minimum restoration time
that reflects the time to estimate which line was damaged and to ensemble
and dispatch a restoration crew, before the restoration itself can
take place.  This model is not intended to be accurate, its parameters
were not fit to any utility time-to-repair data, and was chosen for
its simple parametrization.
A more sophisticated estimated repair time, with a large number of
parameters describing the type of component, its voltage, and
environmental, temporal, and utility stress conditions is also an
option implemented in our model, but which was not used for the present
simulations.
 At the end of the restoration period for a damaged line the utility
has a number of options for reconnecting the line to the power
grid. These options are described in detail in the next subsection.

\subsection{Utility response model}

One of the goals of this modeling approach is to describe the operator's
response to different contingencies, to estimate the optimal operator's
response, and to evaluate how a suboptimal response impacts the risk of
large blackout events. The operator tries to minimize this risk by optimizing
his response in a game against ``nature" which  performs random line failures or,
more generally, random component failures, that can induce one or
multiple line overloads. The random line (component) failure events are called type
3 events. When such an event happens, the operator has the option to 
respond with either deterministic or chance moves. For example, when a line overload 
is produced the operator has the option to either:

\begin{enumerate}

\item With probability $p_1$ shut down the line and protect it against failure and damage.
We call this response a type 1 event.

\item With probability $p_2$, reflecting an erroneous estimation of
 the transmission line state, do nothing and let the line reach its
 critical temperature. We call this ``response'' a type 2 event.

\item With probability $p_3$ perform a partial generation dispatch and
load shedding that can alleviate the overload.

\end{enumerate}
\noindent 
Here the probabilities are such that $p_1+p_2+p_3 = 1$.

Unlike type 1 events, in which the line is not damaged, for type 2
 events the line is damaged and a random restoration time necessary to
 fix the line must pass before it can be restored in service. The
 restoration event 
is called a type 0
 event.  When reconnecting lines, the operator first determines if
 there are islands in the system, which may be produced as lines are
 severed during the evolution of the event. If this is the case, it
 further determines if the reconnected lines join together two, or
 more, islands in the system.  When this happens, we {\it virtually} restore the
 load demand to its initial value before the start of the reconnecting event.
 Here, the idea is that line restoration can fully recover the normal
 operating state of the system and, therefore, fully serve the loads.
 This reflects the fact that operator's actions before the present
 restoration time might have shed loads in order to alleviate some
 overloaded lines observed in the system. 
It is also possible that, due
 to insufficient generation capacity, 
load shed was
 required by the constraint of power balance within an island.

After these virtual restorations and after solving for new power
 flows, the operator checks for the presence of overloaded lines. If
 there are no overloads, the line and load reconnection was successful
 and the operator proceeds with performing these operations by turning
 the virtual restorations into real events.  If there are overloads
 the operator has a couple of options, for which we assign different
 probabilities ($p_4+p_5 = 1$):

\begin{enumerate}


\item with probability $p_4$, reconnects lines even though this process can induce overloads,
of the lines themselves (most probably) or somewhere else in the system;

\item with probability $p_5$, performs another partial generation dispatch and load shedding
that can alleviate the overload.
\end{enumerate}

In order to implement the partial generation dispatch and load shed  algorithm 
we have followed reference~\cite{Ca02} and 
formulated the operator's response as an optimization problem in which
we solve the DC power flow equations  while minimizing the following cost function:
\begin{equation}
C = \sum_{i \in \mathcal{G}} |P_i - P_{i0}| + W \sum_{j \in \mathcal{L}} (P_j - P_{j0})\, .
\end{equation}

\noindent In this equation $P_{i0}$ and $P_{j0}$ are generator and
load values, respectively, before turning on generator dispatch and
load shed, while $P_{i}$ and $P_{j}$ are generator and load values
after the overloads are removed.  The cost function was chosen to
minimize 
a trade-off between
the change in generation (first summation over generator
nodes $\mathcal{G}$) and the load shed (second summation over load
nodes $\mathcal{L}$) necessary to eliminate the overloads.  We assume
that the cost of adjusting generators is the same and that loads share
the same priority to be served. In order to force generator dispatch
first, and simultaneously minimize the load shed in a contingency, we
set a high price for load shed by choosing $W=100$ as in \cite{Ca02}.
While restoring the loads after each contingency seems natural,
generation restoration reflects the idea that the generation
distribution is optimal from an economic view point and, for this
reason, we would like to also restore the state of the
generators. Even though the absolute value of the generation shift
appears in the optimization problem, this optimization problem can be 
solved using linear programing techniques, following the method introduced in~\cite{Do01}.

The minimization of the cost function is performed subject to the following constraints:
\begin{enumerate}
\item forcing an upper limit on the generator power: $0 \le P_i \le P_i^{max}, i \in \mathcal{G}$;
\item forcing the loads not to generate power: $P_{j0} \le P_j \le 0, j \in \mathcal{L} $;
\item forcing the power flow through the line within a $\alpha$ fraction of the line ratings: $ - \alpha P_l^{max} \le 
P_l \le \alpha P_l^{max}$;
\item forcing the  total power generated to exactly balance the total load demand: 
$\sum_{i \in \mathcal{G} \cup \mathcal{L}} P_i = 0$.
\end{enumerate}
We have introduced here a line overload parameter $\alpha$ which allows us to further 
represent either a risk averse response, when $\alpha < 1$, or a risk taking response $\alpha > 1$.

It is obvious that different choices for these probabilities can implement a
large variety of operator's actions.  For example, by choosing $p_3 =1$, 
we dispatch with probability one and we optimize the operator's
response by completely eliminating line overloading events, assuming
that this is our goal.  
Because we can never eliminate the random
line failure events, it is possible that this response 
will not guarantee
a long term optimal response in reducing the average cost of cascading
events.  By always shedding loads to eliminate overloads, this
response might produce numerous small cascades, whose added cost
could be quite large. Therefore, it may be possible, as we discuss in
Section~\ref{optimal_response}, to trade off the cost of small events
versus an increased probability of generating larger, but less
frequent, cascades, by not removing some line overloads.

\section{Simulation algorithm}
\label{algorithm}

The model is rich and complex in its possible behavior. In order to
deal with the large variety of events we have time-ordered all events
in an event list.  There are a number of different event types that
can be generated during the evolution of the system. The simulation
begins by determining the time of the first type 3 event for each
transmission line.  The first type 3 event is initiating the event
list.
We have the possibility
to introduce random load perturbations which happen on a time scale that
is much smaller than the characteristic time scale for random failure
events.  
After each new random load configuration
we solve the power flow equations in order to determine the new state
of the system. If these fluctuations introduce overloads, one of the
utility response actions described in Section~\ref{grid_model} is
chosen according to its \emph{a priori} probability.

When a failure event (type 3) is encountered, the line is damaged and
a random restoration time $t_r$ is drawn from the probability
distribution defined in the restoration model.  A restoration event
(type 0) for this line is introduced into the event list at time
$t+t_r$ and a new random failure time for this failed line is
generated.  The next type 3 event is determined by finding the first
type 3 event over all transmission lines and is introduced into the
event list. To keep this list small, we always keep a single type 3
event in the event list, but the type 3 event times for all lines in
the network are stored in a separate vector.

When an overloaded line is present in the system, there is a
probability $p_2$ that the overloaded state of the line will be missed
by the utility operator, due to an erroneous estimation of the state
of the transmission line.  As we know, this defines a type 2 event.
In this case, there is a time delay, $t_c$, until the line reaches the
critical temperature, that corresponds to its rating power, and
fails. If the failure time $t+t_c$ happens \emph{before} any other
event in the event list, we include this event into the event list and
\emph{remove} any other events of type 1 or 2 that follow. Because this
event damages the line, a random restoration time
to fix the line must pass before it can be put back in service.
Therefore, we also compute a restoration time $t_r$, sampled according
to Eq.~(\ref{restoration_time}), and a restoration event corresponding
to this line will be also included in the event list at time
$t+t_c+t_r$.  As we have just remarked, restoration events define type
0 events.  The line can be restored only \emph{after} time
$t+t_c+t_r$, where $t$ is the present simulation time.

 If $t+t_c$ comes \emph{after} any other event, then we do not include
this type 2 event in the event list, because a grid alteration will
happen before it reaches its critical temperature. Indeed, earlier events
may either modify its time of reaching the critical temperature, or
may remove completely the overload and the line will cool down toward
an equilibrium temperature below the critical temperature. For this
reason, we also remove any other events of type 1 or 2 that follow an
earlier type 2 event. We \emph{never} remove type 3 and type 0 events
from the event list.

Alternatively, with probability $p_1$, the operator shuts down
the line in order to prevent its tripping due to overheating. As we have seen this
defines a type 1 event. In this case, it is not necessary to repair the
line, 
which can be set back in service according to the reconnection
strategy described below at any time after the present time $t$. 

No line can be reconnected until a restoration event happens
because, as described in Section~\ref{grid_model}, we assume that
without a line reconnection the state of the system has not improved
in order to prevent overloads in the reconnected lines or, possibly,
somewhere else in the system. 

After each restoration (type 0) event, we choose to reconnect all type
1 lines according to one of the reconnection strategies presented in
detail in Section~\ref{grid_model}.
Sometimes reconnections result in overloads and, as a
result, the utility may choose to remove the reconnected lines in
order to remove these overloads and to restore the system to its old
status before reconnections were attempted. In this case, each restoration event is
written back into the event list as type 1 event and can be
reconnected again at a later time when another reconnection events is
encountered.  For this reason, the event list will contain many type 1
events before the present simulation time $t$, for which reconnection
attempts have failed.

Finally, at the beginning of a new event
at event time $t$, the line temperatures are determined based on the known
temperatures at the previous event (time step), 
$t-dt$, and the $dt$
time elapsed since the previous event. Thus, according to
Eq.~(\ref{temp_evol}), we update the temperature of line $l$ to
\begin{equation}
T_l(t) = e^{-\nu dt}(T_l(t-dt)-T_{el}(P_l(t-dt))) + T_{el}(P_l(t-dt))\, .
\end{equation}
Note that the power flow $P_l(t-dt)$ 
remains
unchanged since the 
previous event time $t-dt$.

\section{Sample Results for
Cascading
Failures}

\label{sample_cascade}

The first application of 
our time-event simulator is to look at
overloaded-line cascading failures. 
This section illustrates a
cascade simulation for a typical choice of the parameters of the model.
In order to simulate this cascade event we only follow the evolution of real power flows and 
neglect the hourly
load demand variations during the evolution of the event. Our goal is
to compare the cost, in terms of load loss, of various mitigating
responses 
in addressing the emerging system
problems. The operator's response ranges from choosing not to respond to
eliminate overloads, to implement suboptimal generation dispatch, to respond with
optimized system correction, i.e. generation dispatch and load shedding, to
eliminate overloads.  The critical algorithms of the cascade
simulation include estimating restoration time of damaged components,
estimating the time to disconnect
for
overloaded lines, 
recovering (when possible) load previously lost during contingency and islanding events,
and attempting to reconnect previously disconnected lines.

\begin{figure}
\centering
\includegraphics[width=3.0in]{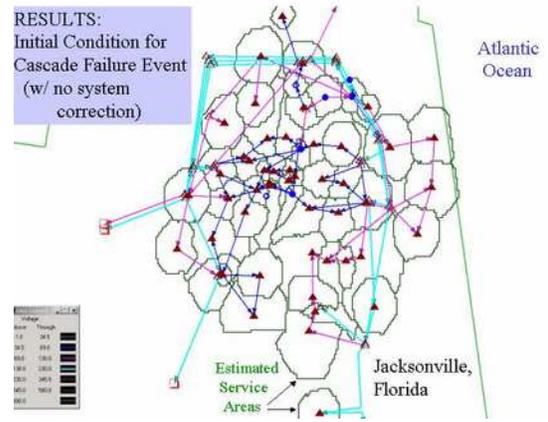}
\caption{Undamaged initial  electrical power model.}
\label{casc1}
\end{figure}

Figure \ref{casc1} shows the
undamaged initial electrical power
model. This transmission system consists of 100 nodes, 133 lines, 24
generators, and 5 interties (boundary conditions representing external
networks not included within the transmission system). Small arrows
indicate the direction of flow of the real power.  The model features
a high-voltage (230-kV) ring (light blue) that carries power to the
lower voltage (138-kV, pink) and (69-kV, dark blue) lines that deliver
power to the distribution network.  The system is quite robust in that
it generates more power than it uses, so it is a net exporter of
power, and it does not depend upon external sources of power.  About 2
GW of power are supplied to local customers.

\begin{figure}
\centering
\includegraphics[width=3.0in]{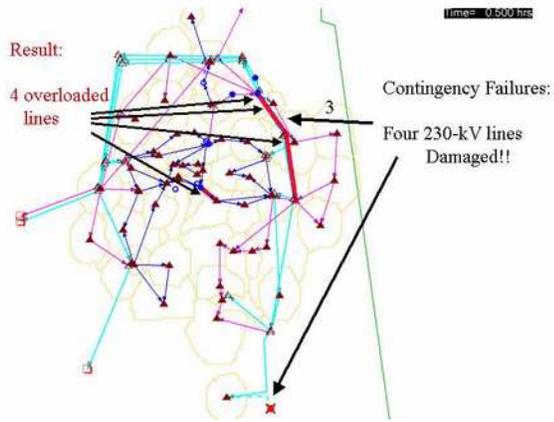}
\caption{Starting contingency for electrical power model.}
\label{casc2}
\end{figure}

\begin{figure}
\centering
\includegraphics[width=3.0in]{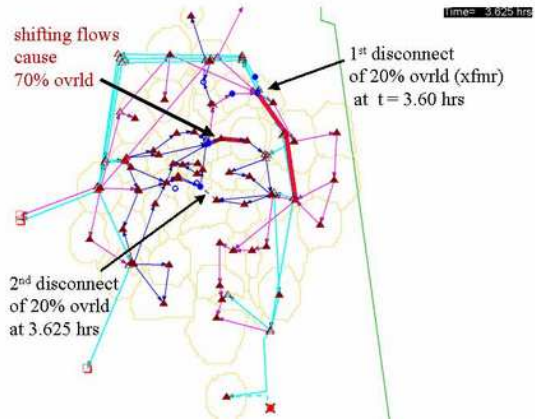}
\caption{Two higher loaded lines are disconnected.}
\label{casc3}
\end{figure}

\begin{figure}
\centering
\includegraphics[width=3.0in]{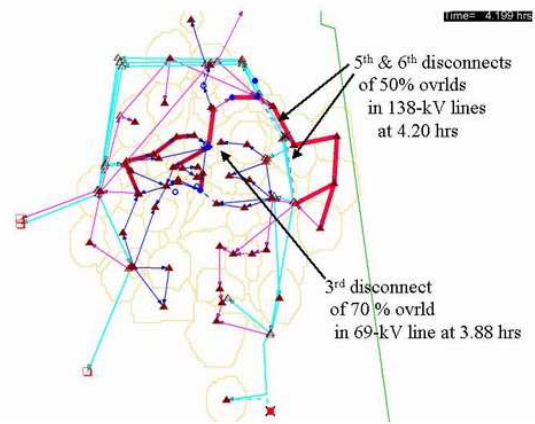}
\caption{Three overloaded lines are disconnected.}
\label{casc4}
\end{figure}

\begin{figure}
\centering
\includegraphics[width=3.0in]{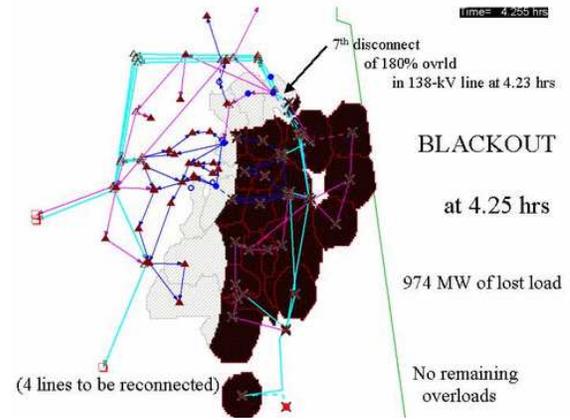}
\caption{Estimated outage area.}
\label{casc5}
\end{figure}

Figure \ref{casc2} shows the starting contingency where four 230-kV lines are damaged 
and lost from service.  This contingency was chosen to sever the high capacity 
transmission paths between the largest generators in the northern section of 
the model from the southeast section of the model that has loads and no 
generation.  This contingency creates four overloaded lines, two at about 
120\% loading and two at 108\%.  The two higher loaded lines are disconnected 
3.6 hrs after the initial failures (see Figure \ref{casc3}).  This shifts the power to
create larger overload of 
70\% and 50\% that are disconnected at 3.8 and 4.2 hrs (see Figure \ref{casc4}).  
This creates a situation where only one (overloaded) path remains to supply 
power to the southeast part of the model.  This last line is disconnected (to 
protect it) at t = 4.23 hrs.  This disconnection creates a black 
electrical island (i.e., no power can reach it).  The estimated outage area 
is shown in Figure \ref{casc5}.  The electrical company loses 974 MW of customer load at 
this time.

After the blackout begins, there are no overloads in the system.  Also several 
of the previously disconnected lines are returned to service, but none of 
those lines connect the black island to the working island.  It is not until 
hour 25.5, when the first one of the originally damaged 230-kV lines is 
replaced and restored to service, that the black island gets reconnected to 
the working network and all loads are recovered (with no overloads in the 
network).  This ends the effect of the cascade event.

The result of this cascade example is summarized in Table~\ref{cont_results}.  In
the discussion above, if the utility 
opts to do nothing after the initial four-line contingency, the relatively 
small overloads that result are eventually automatically disconnected 
to prevent them from being damaged, and the event cascades to cause a blackout 
affecting almost a GW of customer load.  If, instead, the utility immediately 
sheds load after the initial event, all overloads are eliminated by shedding 
108 MW.  If one optimally sheds load and simultaneously adjusts the generation 
within the utility, then only 73 MW need be shed.  The bottom line is that 
under the assumption of constant demand, choosing not to respond to eliminate 
overloads in the system can result in cascading failures where much larger 
loss of customer load can result.  
If we were also simulating the 
hourly variation in demand, the quantitative and possibly the qualitative 
conclusions 
could vary depending upon whether the initial contingency occurred 
during a period of demand growth or demand shrinkage.  Such load-variation 
effects can be modeled using 
the new time-event simulator.

\begin{table}
\renewcommand{\arraystretch}{1.3}
\caption{Results Summary of four 230-kV-Line contingency}
\label{cont_results}
\begin{center}
\begin{tabular}{|c||c||c||c|}
\hline
Approach & Load Cost  & Outage Duration & Effect\\
        & (MW)       &   (hrs)         &       \\
\hline
\hline
do nothing     & 974 & 21.3 & cascade    \\
               &     &      & blackout \\
\hline
shed load      & 108 & 25.0 & load shed          \\
 for overloads &     &      & waiting for repairs \\     
\hline               
shed load       &   73  & 25.0 & load shed  \\
and dispatch      &     &      &  waiting for repairs\\
\hline
\end{tabular}
\end{center}
\end{table}

\section{Optimal response}
\label{optimal_response}

In this Section we discuss some of the features of the stochastic
model when the utility response to line overloads is represented by
the generation dispatch and load shed optimization algorithm. We
choose to vary the line overload parameter $\alpha$ in order to
describe a smooth transition from a risk averse $\alpha < 1$ to a risk
taken response for $\alpha > 1$.

We want to test here two different optimization strategies against the
response 
$\alpha=1$.  
The choice $\alpha=1$ defines the normal operating
 response represented by generation dispatch and load shedding to
 restore line flows at their critical value (line ratings). This should be  compared
 to $\alpha < 1$ case when the operator prevents the lines from even
 reaching their thermal ratings, or $\alpha > 1$ when the operator
 decides to respond when the line flows reach a threshold higher than
 the line ratings. In the first case, $\alpha < 1$,
  we trade off an increased load shed versus a smaller risk of
 generating large cascades. In the second case, $\alpha > 1$, we trade
 off a smaller load shed versus an increased risk of generating large
 cascades. The question we ask is the following: Are there any model
 parameters for which one of this choices performs better than the
 optimization algorithm for $\alpha =1 $?  While we do not have as yet
 a complete answer to this question, we present here the results of our
 numerical experiments for the following model parameters: $\lambda_f$
 = 0.0001, $c=3 h $, $\lambda_r = 0.2 h^{-1} $ and $p_3 =1$.  We have run our
 experiments for the power grid model of 100 nodes, 24 generators, and 5
 interties, that we have already introduced in Section~\ref{sample_cascade}.
 In order to compare the model responses for different $\alpha$ values
 we have estimated the cost of  cascades per unit time, $C_1$, and the cost
 of cascades per event, $C_2$. The cost of a cascade, in MW hour, is defined as follows:
\begin{equation}
C = \int_{t_{s}}^{t_{e}} \Delta P (t)dT\, ,
\end{equation}
where $t_s$ is the time when cascade event starts, i.~e.~ the power
served is less than the power demand, $t_e$ is the time when the
cascade events ends with full service of the power demand, and $
\Delta P (t)$ is the power shed at time $t$ during the cascade, $t_s
\le t \le t_e$. In fact, the two cost functions we use can be defined
as
\begin{eqnarray}
C_1 &=& \frac{\int_{t_i}^{t_f} \Delta P (t)dT }{t_f - t_i}, \\
C_2 &=& \frac{\int_{t_i}^{t_f} \Delta P (t)dT }{N} \, ,
\end{eqnarray}
where $t_i$ is the starting time of the simulation, $t_f$ is the
ending time of the simulation, while $N$ is the number of cascade
events during the simulation time. Of course, $C_2$ is also the
average cost of a cascade. Even though we call all events that shed
power in the system ``cascades'', not all events are cascading events
in which a triggering failure produces a sequence of secondary
failures that lead to blackout of a large area of the grid as
presented in Section ~\ref{sample_cascade}. An exact characterization of the
events, including their scaling properties will be presented
elsewhere.

\begin{table}
\renewcommand{\arraystretch}{1.3}
\caption{Comparison of utility responses with various optimization risk factors}
\label{opt_results}
\begin{center}
\begin{tabular}{|c||c||c||c||c||c||c||c||c|}
\hline
$\alpha$ & 0.90 & 0.95 & 1.00 & 1.10 & 1.20 & 1.30 & 1.40 \\
\hline
$C_1$    & 10.70 & 10.93 & 0.35 & 1.78 & 2.02 & 3.04 & 7.27 \\
\hline
$ C_2 $  & 4026 & 4193  &  178 &  937 & 1032  & 1498 & 2966 \\
\hline
\end{tabular}
\end{center}
\end{table}

The results of our numerical experiments are summarized in
Table~\ref{opt_results}.  For the set of model parameters used in our
experiments, the best response strategy is to implement the generation
dispatch and load shed optimization with parameter $\alpha = 1$, but
we should remark that a global $\alpha$ parameter is probably not the
most efficient
implementation of this optimization idea. We expect that
choosing a line dependent $\alpha$ value, that takes into account the
importance of each line flow to the overall power transport in the
system, could provide a more successful optimization strategy. For example,
we can choose $\alpha=1$ for lines that carry the backbone of the
power flow, and which will probably generate large flow
redistributions for either $\alpha < 1$ or $\alpha > 1$, and a smaller
or larger $\alpha$ value for the rest of the lines. A possible
objection against this strategy can be the fact that the optimization
approach proposed leads to damage 
to equipment, but it is useful to
remember that August $14^{th}$, 2003 blackout cost billions of dollars
in in economic losses  but caused negligible equipment damage \cite{Aug03}.
It is possible then that
equipment damage that can significantly decrease the social cost of
blackout events is a feasible prevention strategy.  Moreover, one
could conceivably extended this approach to other parameters or to other
optimization strategies that can be used to assess vulnerabilities
and to allocate protection devices and preventive maintenance
responses, as described in \cite{Pe01}. The model can also be
generalized to replace the unique power grid operator with a network
of distributed, autonomous agents who share local information in order
to coordinate their local responses to finding good global optimization
solution as described in~\cite{Hi05}. A decentralized approach may also
benefit from the information provided by the structure of the underlying
network of flows, as proposed in~\cite{n4}.

\section{Conclusion}

As more vulnerable networks emerge, due to the introduction of
deregulated energy markets, the demand for a more 
reliable
representation of the networks in order to correctly assess the
operational security level of the transmission system will
increase. At the same time, as the complexity of operating the grid
grows,
modeling the operator's response to these challenging demands
becomes increasingly critical and should match in sophistication the
modeling of the grid itself. For these reasons, we have presented a model
that 
attempts to provide a 
comprehensive
representation of the complex behavior of 
both the grid dynamics under random perturbations 
and
the operator's response to 
the contingency events.  This response is not
always optimal due to either failure of the state estimation system or
due to the incorrect subjective assessment of the severity associated
with these events.

Furthermore, we have cast the optimization of the operator's response into
the choice of the optimal strategy for mitigating the impact of random
component failure events and, possibly, for controlling blackouts. A first
example was to test a generation dispatch and load shed algorithm for
a range of risk factors that were trying to balance the risk of load
shed versus the risk of generating large cascading events.  This
simple strategy space can be easily extended and we have suggested a few
possible generalizations that are already the subject of intense research.

Moreover, the model can be used  to test the conjecture that power grids
operate close to a ``critical'' point as suggested by recent analysis
of power system disturbance data~\cite{Ch01,Ca00}. If this is 
confirmed,
some aspects of
the response of the system to random perturbations may have an
universal character. In this case, the dynamics of the stochastic
model for different
parameters should exhibit the conjectured
universal behavior. An analysis of this conjecture is  the
subject of our current research.

\section*{Acknowledgment}

The authors thank 
Ian Dobson for his
careful reading of the manuscript and for
many detailed and useful suggestions that significantly improved the quality of
our presentation.
Part of this work was carried out under the auspices of the National Nuclear 
Security Administration of the U.S. Department of Energy at Los Alamos 
National Laboratory under Contract No. DE-AC52-06NA25396.


\begin{thebibliography}{1}


\bibitem{Ba99} K.~Bae and J.~S.~Thorp, 
``A stochastic study of hidden failures in power system protection'', 
\emph{Decision Support Systems}, vol.~24, pp.~259-268, 1999.

\bibitem{Ca02} B.~A.~Carreras, V.~E.~Lynch, I.~Dobson, and D.~E.~Newman, 
``Critical Points and Transitions in an Electric Power Transmission Model for Cascading Failure Blackouts'', 
\emph{Chaos}, vol.~12, no.~4, pp.~985-994, 2002.

\bibitem{Ca04} B.~A.~Carreras, V.~E.~Lynch, I.~Dobson, and D.~E.~Newman, 
``Complex Dynamics of Blackouts in Power Transmission Systems'', 
\emph{Chaos}, vol.~14, no.~3, pp.~643-652, 2004.

\bibitem{Ca01} B.~A.~Carreras, V.~E.~Lynch, M.~L.~Sachtjen, I.~Dobson, and D.~E.~Newman, 
\emph{Modeling Blackout Dynamics in Power Transmission Networks}, 
Hawaii International Conference on System Sciences, January 3-6, 2001, Maui, Hawaii.

\bibitem{Ca00} B.~A.~Carreras, D.~E.~Newman, I.~Dobson, and A.~B.~Poole,
``Evidence for Self-Organized Criticality in a Time Series of Electric Power System Blackouts'',
\emph{IEEE Transactions on Circuits and Systems I}, vol.~51, no.~9, pp.~1733-1740, 2004.

\bibitem{CaJa59} H.~S.~Carslaw  and J.~C.~Jaeger, 
\emph{ Conduction of Heat in Solids},  
2nd~ed.\hskip 1em plus  0.5em minus 0.4em\relax Clarendon Press, Oxford, 1959.

\bibitem{n3}
D.~P.~Chassin and C.~Posse,
``Evaluating North American Electric Grid Reliability Using the Barab\'asi-Albert Network Model", 
\emph{Physica A}, vol.~55, no.~2-4, pp.~667-677, 2005.

\bibitem{Ch02} J.~Chen and J.~S.~Thorp, 
\emph{A Reliability Study of Transmission System Protection via a Hidden Failure DC Load Flow Model}, 
IEEE Fifth International Conference on Power System Management and Control, pp.~384-389, April 17-19, 2002.

\bibitem{Ch01} J.~Chen, J.~S.~Thorp, and M.~ Parashar, 
\emph{Analysis of Power System disturbance Data}, 
Hawaii International Conference on System Sciences, January 3-6, 2001, Maui, Hawaii.


\bibitem{Do01} I.~Dobson, B.~A.~Carreras, V.~E.~Lynch, and D.~E.~Newman, 
\emph{An Initial Model for Complex Dynamics in Electric Power System Blackouts}, 
Hawaii International Conference on System Sciences, January 3-6, 2001, Maui, Hawaii.

\bibitem{Do04} I.~Dobson, B.~A.~Carreras, V.~E.~Lynch, and D.~E.~Newman,
\emph{Complex System Analysis of Series of Blackouts: Cascading Failure, Criticality, and Self-organization}, 
Bulk Power Systems Dynamics and Control - VI, August 22-27, 2004, Cortina d'Ampezzo, Italy.

\bibitem{n6}
K.~I.~Goh, D.~S.~Lee, B.~Kahng, and D.~Kim,
``Sandpile on Scale-Free Networks",
\emph{Physical Review Letters}, vol.~91, no.~14, art.~no.~148701, 2003.

\bibitem{GrSt94} J.~J.~Grainger and W.~D.~Stevenson, Jr.~, 
\emph{Power System Analysis},
2nd~ed.\hskip 1em plus 0.5em minus 0.4em\relax McGraw-Hill, New York, 1994.

\bibitem{Ha04} R.~C.~Hardiman, M.~Kumbale, and Y.~V.~Makarov,
\emph{Multiscenario Cascading Failure Analysis Using TRELSS},
Quality and Security of Electric Power Delivery Systems, CIGRE/IEEE PES International Symposium, pp.~176-180,
October 8-10, 2003.

\bibitem{Hi05} P.~ Hines, H.~ Liao, D.~ Jia, and S.~ Talukdar,
\emph{Autonomous Agents and Cooperation for the Control of Cascading Failures in Electric Grids}, 
Proceedings of the IEEE Conference on Networking, Sensing, and Control, March 2005, Tucson, Arizona.

\bibitem{n7}
P.~Holme, and B.~J.~Kim,
``Vertex Overload Breakdown in Evolving Networks",
\emph{Physical Review E}, vol.~65, no.~6, art.~no.~066109, 2002.

\bibitem{IlSk00} M.~Ilic and P.~Skantze, 
``Electric Power Systems Operation by Decision and Control'', 
\emph{IEEE Control Systems Magazine}, vol.~20, no.~4, pp.~25-39, 2000.

\bibitem{n1}
R.~Kinney, P.~Crucitti, R.~Albert, and V.~Latora,
``Modeling Cascading Failures in the North American Power Grid",
\emph{European Physical Journal B}, vol.~46, no.~1, pp.~101-107, 2005.

\bibitem{n5}
Y.~Moreno, R.~Pastor-Satorras, A.~Vazquez, and A.~Vespignani,
``Critical Load and Congestion Instabilities in Scale-Free Networks", 
\emph{Europhysics Letters}, vol.~62, no.~2, pp.~292-298, 2003.

\bibitem{n4}
A.~E.~Motter,
``Cascade Control and Defense in Complex Networks", 
\emph{Physical Review Letters}, vol.~93, no.~9, art.~no.~098701, 2004.

\bibitem{NERC96} NERC (North American Electric Reliability council),
1996 system disturbances, (Available from NERC, Princeton Forrestal Village, 
116-390 Village Boulevard, Princeton, New Jersey 08540-5731), 2002.

\bibitem{Ni03} M.~Ni,J.~D.~McCalley, V.~Vittal, and T.~Tayyib,
``On-line Risk-Based Security Assessment'', 
\emph{IEEE Transactions on Power Systems}, vol.~18, no.~1, pp.~258-265, 2003.


\bibitem{Pa84} A.~Papoulis, 
\emph{Probability, Random Variables, and Stochastic Processes}, 
2nd~ed.\hskip 1em plus  0.5em minus 0.4em\relax McGraw-Hill, New York, 1984.

\bibitem{Pe01} D.~L.~Pepyne, C.~G.~Panayiotou, C.~G.Cassandras, and Y.-C.~ Ho, 
\emph{Vulnerability Assessment and Allocation of Protection Resources in Power Systems},
Proceedings of the American Control Conference, June 25-27, 2001, Arlington, Virginia.

\bibitem{numrec} W.~H.~Press, S.~A.~Teukolsky, W.~T.~Vetterling, and B.~P.~Flannery, 
\emph{Numerical Recipes in C}, 
2nd~ed.\hskip 1em plus 0.5em minus 0.4em\relax Cambridge University Press, 1992.

\bibitem{Ri02} M.~A.~Rios, D.~S.~Kirschen, D.~Jayaweera, D.~P.~Nedic, and R.~N.Allan, 
``Value of Security: Modeling Time-Dependent Phenomena and Weather Conditions'', 
\emph{IEEE Transactions on Power Systems}, vol.~17, no.~3, pp.~543-548, 2002.

\bibitem{Aug03} U.~S.-Canada Power System Outage Task Force, Final Report on the 
August 14th Blackout in the United States and Canada. United States Department of 
Energy and National Resources Canada, April 2004.

\bibitem{Sa00} M.~L.~Sachtjen, B.~A.~Carreras, and V.~E.~Lynch,
``Disturbances in Power Transmission System",
\emph{Physical Review E}, vol.~61, no.~ 5, pp.~4877-4882, 2000.

\bibitem{n2}
A.~Scire, I.~Tuval, and V.~M.~Eguiluz,
``Dynamic Modeling of the Electric Transportation Network",
\emph{Europhysics Letters}, vol.~71, no.~2, pp.~318-324, 2005.

\bibitem{In05} I.~ Steinwart, D.~ Hush, and C.~ Scovel, 
``A Classification Framework for Anomaly Detection'', 
\emph{ Journal of Machine Learning Research}, vol.~6, pp.~211-232, 2005.

\bibitem{Th98} J.~S.~Thorp, A.~G.~Phadke, S.~H.~Horowitz, and S.~Tamronglak, 
``Anatomy of Power System Disturbances: Importance Sampling'', 
\emph{Electric Power and Energy Systems}, vol.~20, no.~2, pp.~ 147-152, 1998.

\bibitem{n8}
D.~J.~Watts, 
``A Simple Model of Global Cascades on Random Networks",
\emph{Proceedings of the National Academy of Sciences U.~S.~A}, vol.~99, no.~9, pp.~5766-5771, 2002.

\bibitem{We05} K.~A.~Werley, \emph{The Power System Analyzer (PSA) Suite of Numerical Tools}, 
Los Alamos National Laboratory report LA-UR-03-8268,  pp.~1-61, July 2005.

\bibitem{WoWo96} A.~J.~Wood and B.~F.~Wollenberg, 
\emph{Power Generation, Operation, and Control}, 
2nd~ed.\hskip 1em plus 0.5em minus 0.4em\relax John Wiley \& Sons, New York, 1996.


\end{thebibliography}
\end{document}